%% file: MainFile.tex
\documentclass[conference]{IEEEtran}
\usepackage{graphicx,multirow}
\usepackage[vlined,linesnumbered,ruled,boxed]{algorithm2e}

 \usepackage{layout}
 \usepackage{url}
\usepackage{wrapfig}
\usepackage{bibentry}

\usepackage{tabularx} %table
\usepackage{algorithmic}
\usepackage{amssymb}
\usepackage{bbold}

\nobibliography*

\usepackage{wrapfig,color,caption,subcaption,balance,multirow}

\usepackage{adjustbox,tabu}

\usepackage[vlined,linesnumbered,ruled,boxed]{algorithm2e}
 \usepackage{amsmath}

 \usepackage{caption}
  \usepackage{subcaption}

\usepackage{bibentry}

\makeatletter
\def\thickhline{%
  \noalign{\ifnum0=`}\fi\hrule \@height \thickarrayrulewidth \futurelet
   \reserved@a\@xthickhline}
\def\@xthickhline{\ifx\reserved@a\thickhline
               \vskip\doublerulesep
               \vskip-\thickarrayrulewidth
             \fi
      \ifnum0=`{\fi}}
\makeatother

\newlength{\thickarrayrulewidth}
\nobibliography*

 \usepackage{graphicx,multirow}

\usepackage[colorlinks, citecolor=blue]{hyperref}
\usepackage{diagbox}

\newcommand{\eg}{{e.g., }}
\newcommand{\etal}{{et~al. }}
\newcommand{\ie}{{i.e., }}

\newcommand{\name}{Edge-MultiAI\xspace}

\newcommand{\comments}[1]{}
\usepackage{cleveref}
\newcommand\hl{\bgroup\markoverwith
  {\textcolor{yellow}{\rule[-.5ex]{2pt}{2.5ex}}}\ULon}

\def\BibTeX{{\rm B\kern-.05em{\sc i\kern-.025em b}\kern-.08em
    T\kern-.1667em\lower.7ex\hbox{E}\kern-.125emX}}

\begin{document}
\title{\name: Multi-Tenancy of Latency-Sensitive Deep Learning Applications on Edge\\
  {\normalfont\large 
  SM Zobaed\IEEEauthorrefmark{1}, Ali Mokhtari\IEEEauthorrefmark{1}, Jaya Prakash Champati\IEEEauthorrefmark{2}, Mathieu Kourouma\IEEEauthorrefmark{3}, Mohsen Amini Salehi\IEEEauthorrefmark{1} %
  }\\[-1.5ex]
}
\author{
    \IEEEauthorblockA{%
        \IEEEauthorrefmark{1}\{sm.zobaed1, ali.mokhtari1, amini\}@louisiana.edu\\
        HPCC Lab, School of Computing and Informatics\\
        University of Louisiana at Lafayette,
        LA, USA
    }
    \and
    \IEEEauthorblockA{
        \IEEEauthorrefmark{2}jaya.champati@imdea.org  \\
        IMDEA Networks Institute\\
        Madrid, Spain
    }
    \and 
        \IEEEauthorblockA{
        \IEEEauthorrefmark{3}mathieu\_kourouma@subr.edu  \\
        Southern University and A\&M College\\
        LA, USA
    }
   
}

%\title{\name: Enabling Low-Latency across Multi-Tenant Deep Learning Applications on Edge Systems}
%\author{First Author\inst{1}\orcidID{0000-1111-2222-3333} \and
%Second Author\inst{2,3}\orcidID{1111-2222-3333-4444} \and
%Third Author\inst{3}\orcidID{2222--3333-4444-5555}}
%

%\institute{Princeton University, Princeton NJ 08544, USA \and
%Springer Heidelberg, Tiergartenstr. 17, 69121 Heidelberg, Germany
\maketitle              % typeset the header of the contribution
\linespread{.95}
\input{\paperfolder/Sources/sec-abs}

\input{\paperfolder/Sources/sec-intro}

\input{\paperfolder/Sources/sec-rw}

% %\input{Sources/sec-implementation}
% % \input{\paperfolder/Sources/sec-agt}
% %\input{\paperfolder/Sources/sec-index}
% %\input{\paperfolder/Sources/sec-contgcy}
\input{\paperfolder/Sources/sec-prop}

\input{\paperfolder/Sources/sec-evaluation}

\input{\paperfolder/Sources/sec-conclusion}

% \bibliography{references}
%

\bibliographystyle{IEEEtran}
\balance
\linespread{.72}
\bibliography{references}

\end{document}

%% file: Sources/sec-abs.tex
\begin{abstract}
    Smart IoT-based systems often desire continuous execution of multiple latency-sensitive Deep Learning (DL) applications. The edge servers serve as the cornerstone of such IoT-based systems, however, their resource limitations hamper the continuous execution of multiple (multi-tenant) DL applications. The challenge is that, DL applications function based on bulky ``neural network (NN) models'' that cannot be simultaneously maintained in the limited memory space of the edge. Accordingly, the main contribution of this research is to overcome the memory contention challenge, thereby, meeting the latency constraints of the DL applications without compromising their inference accuracy. We propose an efficient NN model management framework, called \name, that ushers the NN models of the DL applications into the edge memory such that the degree of multi-tenancy and the number of warm-starts are maximized. \name~leverages NN model compression techniques, such as model quantization, and dynamically loads NN models for DL applications to stimulate multi-tenancy on the edge server. % without compromising the latency and accuracy of the involved applications. 
    We also devise a model management heuristic for \name, called \emph{iWS-BFE}, that functions based on the Bayesian theory to predict the inference requests for multi-tenant applications, and uses it to choose the appropriate NN models for loading, hence, increasing the number of warm-start inferences. We evaluate the efficacy and robustness of \name under various configurations. The results reveal that \name can stimulate the degree of multi-tenancy on the edge by at least $2\times$ and increase the number of warm-starts by $\approx 60\%$ without any major loss on the inference accuracy of the applications. %In addition, we provide comparisons of the proposed policies in terms of robustness against uncertainties.        
\end{abstract}
\begin{IEEEkeywords} Edge computing, Multi-tenancy, Deep learning applications, Memory contention. 
\end{IEEEkeywords}

%% file: Sources/sec-intro.tex
\section{Introduction}\label{sec:intro}
\subsection{Motivation and Overview}
\begin{figure}
    \centering
    \includegraphics[width=.43\textwidth]{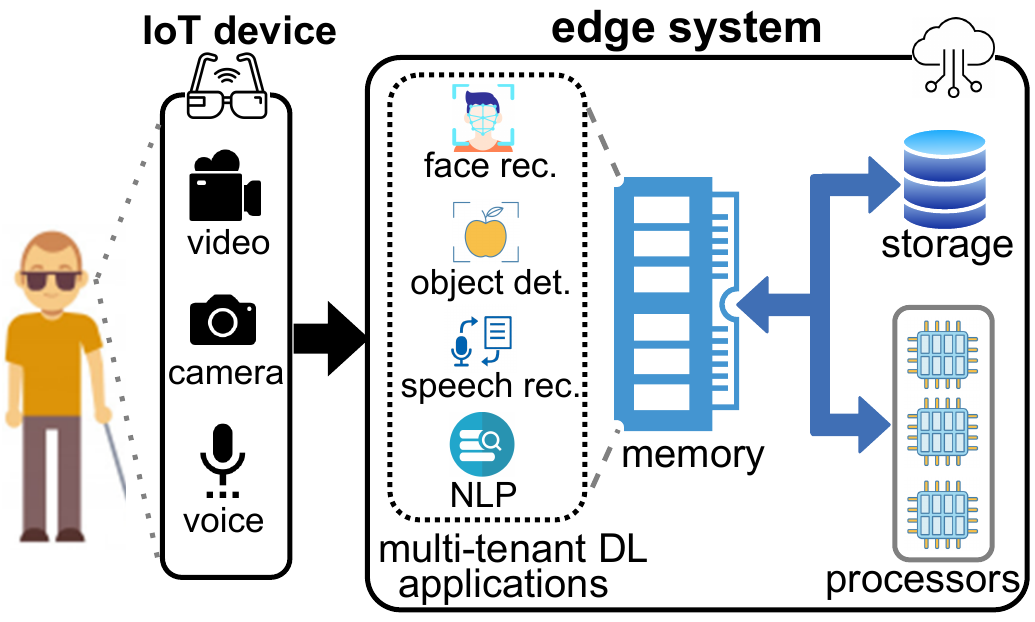}
    \caption{Bird-eye view of SmartSight, an IoT-based system that continuously receives various inputs from the smartglass (IoT device) sensors, and processes them via multi-tenant DL applications running on the edge server.}
\label{fig:high_level}
\vspace{-.3 cm}
\end{figure}

With the expeditious advances of smart IoT-based systems, they are becoming an indispensable part of our day-to-day life. Such systems often provide their users with multiple latency-sensitive Deep Learning (DL) services, such as object detection, face recognition, and motion capture, that can collectively unlock use cases to improve the human's quality of life. %especially in the context of assistive technology~\cite{felare2022} to mitigate the sufferings of disabilities. 
An exemplar use case of such IoT-based systems is SmartSight~\cite{felare2022}, illustrated in Figure~\ref{fig:high_level}, that aims at providing ambient perception for the blind and visually impaired people.
The system operates based on a smartglass (IoT device) and a companion edge server (\eg smartphone). The smartglass continuously captures the inputs via its sensors (\eg camera and microphone) and requests the edge server to process DL-based applications, such as object detection to identify obstacles; face recognition to identify acquainted people; speech recognition, and NLP to understand and react to the user’s commands. To make SmartSight usable, the edge server has to continuously execute multiple (a.k.a. \emph{multi-tenant}) DL application to process incoming requests with low-latency and high accuracy. It is noteworthy that, although cloud datacenters can mitigate the inherent resource limitations of the edge, due to the network latency overhead and data confidentiality~\cite{deng2020edge,zobaed2021saed,hussain2020analyzing}, offloading the latency-sensitive service requests to the cloud is not a tractable approach in many use cases. %execution are the critical metrics for the \emph{usability} of SmartSight. 
%That is, the edge server has to support \emph{multi-tenant} execution of DL applications to process simultaneous inputs from the IoT device with low-latency, while maintaining the accuracy of the DL services.

DL applications utilize bulky Neural Network (NN) models at their kernel to infer on the inputs received from the sensors. The NN models have to be kept in memory to enable low-latency (a.k.a. \emph{warm-start}~\cite{dang2017dtrust}) inference operations. Otherwise, because the NN model size is often huge, loading it into the memory in an on-demand manner (a.k.a. \emph{cold-start}) is counterproductive and affects the latency constraint of the DL applications. As the edge servers naturally have a limited memory size (\eg 4 GB in the case of Jetson Nano \cite{nvidia}), multi-tenant execution of DL applications on them leads to a memory contention challenge across the processes~\cite{deng2020edge,gholami2021survey}.
%Assistive technology in first paragraph as example of iot-based systems. ... DOnt keep anything related to problem in the 1st paragraph.
%second para-? problem
%third-> smart sight (use case.motivation)
%An efficient way to handle the user demands within the latency constraint at the edge tier is to utilize edge systems that enable execution of multi-tenant latency-sensitive DL applications~\cite{deng2020edge} simultaneously. Towards this, IoT-based systems have been emerged where IoT devices are connected to an edge server (\eg smartphone, tablet \etc) that executes the application requests incoming from IoT devices on a concurrent basis~\cite{zhou2019edge}. 
%A high-level architectural overview of such a system is illustrated in Figure~\ref{fig:high_level}.
Accordingly, the main challenge of this study is to resolve the memory contention across multi-tenant DL applications without compromising their latency and accuracy constraints. 
%maximize the accuracy of multi-tenant latency-sensitive DL applications within a memory-limited edge server?}
%\begin{wraptable}{r}{8.5cm}

\begin{table}
\centering
\caption{Load time, inference time, and accuracy of popular NN models individually running
on Samsung Galaxy S20+ as the edge server.}%, S20e and Jetson Nano.}
% \usepackage{multirow}
\resizebox{.95\linewidth}{!}{
\begin{tabular}{|c|c|c|c|c|c|}
\hline

      \diagbox[width=\dimexpr \textwidth/8+2\tabcolsep\relax, height=1cm]{\textbf{NN Models}}
                                                & \begin{tabular}[c]{@{}c@{}}\textbf{Bit}\\  \textbf{Width}\end{tabular} & \multicolumn{1}{l|}{\begin{tabular}[c]{@{}l@{}}\textbf{Size}\\ \textbf{(MB)}\end{tabular}} & \begin{tabular}[c]{@{}c@{}}\textbf{Loading}\\\textbf{Time (ms)}\end{tabular} & \begin{tabular}[c]{@{}c@{}}\textbf{Inference}\\\textbf{Time (ms)}\end{tabular} & \begin{tabular}[c]{@{}c@{}}\textbf{Accu-}\\\textbf{racy (\%)}\end{tabular} \\ \hline \hline
\multirow{2}{*}{InceptionV3}                      & FP32                                                 & 105                                                                      & 650                                                       & 100                                                      & 78.50                                                \\ \cline{2-6} 
                                                  & INT8                                                 & 24                                                                       & 380                                                       & 80                                                       & 77.20                                                \\ \thickhline
\multirow{2}{*}{VGG16}                            & FP32                                                 & 528                                                                      & 820                                                       & 52                                                       & 71.30                                                \\ \cline{2-6} 
                                                  & INT8                                                 & 132                                                                      & 185                                                       & 40                                                       & 70.18                                                \\ \thickhline
\multirow{2}{*}{MobileNetV1}                      & FP32                                                 & 89                                                                       & 600                                                       & 15                                                       & 70.56                                                \\ \cline{2-6} 
                                                  & INT8                                                 & 23                                                                       & 192                                                       & 8                                                        & 65.70                                                \\ \thickhline
\multirow{2}{*}{MobileNetV2}                      & FP32                                                 & 26                                                                       & 110                                                       & 10                                                       & 72.08                                                \\ \cline{2-6} 
                                                  & INT8                                                 & 9                                                                        & 65                                                        & 7.5                                                      & 63.70                                                \\ \thickhline
                                                  
\multirow{2}{*}{MobileNetV3}  & FP32                                                 & 14                                                                       & 80.3                                                      & 7.80                                                     & 74.04                                                \\ \cline{2-6} 
\multicolumn{1}{|l|}{}                             & INT8                                                 & 8                                                                        & 47.45                                                     & 6.21                                                     & 71.32                           \\ \thickhline

\multirow{2}{*}{MobileBERT}  & FP32                                                 & 96                                                                       & 1100                                                      & 62                                                       & 81.23                                                \\ \cline{2-6} 
\multicolumn{1}{|l|}{}                            & INT8                                                 & 26                                                                       & 890                                                       & 40                                                       & 77.08                                                \\ \hline
\end{tabular}
}
\label{tab:model_des}
\vspace{ -.5 cm}
\end{table}

% models are bulky, make them edge friendly.

%It is evident that leveraging a complex model yields better inference quality compared to oversimplified ones but loading a complex model occupies a high amount of memory and affects latency constraints. Hence, edge-friendly lightweight (a.k.a. compressed) models are the need of the hour.
%, \emph{pruning}~\cite{islam2021proximity,anwar2016compact,yang2017designing,yao2017deepiot,yin2020dreaming}, and \emph{knowledge distillation} \cite{hinton2015distilling,gou2021knowledge} have been proposed to reduce the model size for faster inferencing at the edge servers while compromising with accuracy by computing tensors at lower bitwidths (\eg INT8) in lieu of regular floating point precision (\eg FP32) and removing redundant connections of NN.

In the deep learning context, there are techniques based on the idea of approximate computing, such as quantization~\cite{zhou2018adaptive}, that make the model edge-friendly via compressing its NN model, hence, reducing its inference time and accuracy. To understand the impact of such approximations, we conducted a preliminary experiment using a Samsung Galaxy S20+ as the edge server; and five popular DNN models, namely InceptionV3, VGG16, MobileNetV1, MobileNetV2, MobileNetV3, MobileBERT, each one at two quantization (precision) levels, namely FP32 and INT8 bit widths. In Table~\ref{tab:model_des}, we report the average loading time, inference time, and accuracy for their individual executions. We observe that: (A) for all the models, the loading time is 8---17$\times$ more than its inference time; (B) Loading the high-precision model (FP32 bit width) occupies $\approx$3.5$\times$ more memory than the low-precision (INT8 bit width) one; and (C) Loading a low-precision model can reduce the inference accuracy by around 3---6\%. These results demonstrate that the model compression has a considerable potential to mitigate the memory footprint of the DL applications. Moreover, the model loading time invariably dominates the inference time~\cite{samani2022exploring}. Accordingly, our hypothesis is that \emph{the efficient use of model compression and the edge memory can enhance the multi-tenancy and inference time of DL applications without any major loss on their inference accuracy}. 

%Hence, edge-friendly lightweight (a.k.a. compressed) models with practicable prediction accuracy are the need of the hour. 
%For instance, MobileBERT~\cite{sun2020mobilebert} is one of the compressed (a.k.a. low-precision) forms of the original BERT, a widely-used machine learning-based natural language processing (NLP)  \cite{} tool, whose model size is $\approx4.5 \times$ smaller and is $\approx5.5 \times$ faster with $\approx62ms$ inference latency time.

%To enable latency-sensitive DL applications on a multi-tenant edge system, w
We propose each DL application to be equipped with multiple NN models with different precision levels. The low-precision models have a small memory footprint, hence, allowing for a higher multi-tenancy of DL applications with their models loaded into the memory (\ie warm-start inference) that enhances the service latency. However, loading overly low-precision (over-quantized) models to maximize multi-tenancy and warm-start inference is not viable, because it reduces the inference accuracy and renders the multi-tenant DL applications to be futile. On the contrary, loading high-precision (large) NN models on a memory-limited edge system for an indefinite time period unnecessarily occupies an excessive memory space that is detrimental for the multi-tenancy and warm-start inference of other tenants. That is, other tenants face a significant slow down (as noted in Table~\ref{tab:model_des}), because they cannot keep their NN model in memory and have to load it from the storage (\ie cold-start) to perform the inference operation. Therefore, an ideal solution for a multi-tenant edge system should be able to dynamically load a suitable model from the set of models available to the application (a.k.a. model zoo), such that it neither interrupts the execution of other applications, nor causes a cold-start inference for them.%  with the overhead of acceptable accuracy loss.       

%redundant
%Although edge AI over cloud AI offers latency-sensitive execution, running DL applications concurrently on the edge systems while maintaining low latency (\ie model loading and inference time) and high throughput (\ie prediction accuracy) is challenging.
 %Accordingly, the research problem that we consider in this study is: \emph{How to enable latency-sensitive execution of multi-tenant DL applications with maximum accuracy on a memory-constrained edge system?}

\subsection{Problem Statement}
The research question that we investigate is: \emph{how to maximize the number of warm-start inferences for multi-tenant DL applications on edge without compromising the inference accuracy?} The question indicates a trade-off between two objectives: fulfilling the latency constraint of DL applications and maintaining their inference accuracy. The former objective entails having the NN models of DL applications loaded into the memory (\ie warm-start inference), whereas, the latter entails retaining high-precision NN models in the memory. 

For application $A_i \in A$ with $M_i = \{ m^{k}_i \ | \ 1 \le k \le q_i\}$ as its model zoo, let $r_i(t)$ be a Boolean function that represents an inference request for $A_i$ at time $t$ with value 1. 
% For application $A_i \in A$, let $M_i = \{ m^{(k)}_i \ | \ 1 \le k \le s_i\}$ denote its model zoo with $\lvert M_i \rvert = s_i$. Also, the request for $A_i$ at time $t$ is represented by a Boolean function, denoted $r_i(t)$,  with value 1. 
% Then, the request function, denoted by $r(t)$, is defined as $\sum_{i} r_i(t)$. 
Also, let $m^\ast_i \subseteq M_i$ be an NN model of $A_i$ with size of $s^\ast_i$ that is currently loaded in the memory. This means that, for application $A_j$ that does not have any of its NN models currently in the memory, we have $m^\ast_j = \varnothing$ and $s^\ast_j = 0$. Then, $M^\ast = \bigcup_{i=1}^{n} m^\ast_i$ represents the set of currently loaded NN models that occupy $S^\ast = \sum_{i=1}^{n} s^\ast_i$ of the memory space. 
A cold start event for the request arrives at time $t$ for $A_i$, denoted $C_i(M^\ast,t)$ and shown in Equation (\ref{eq:cold_start}), occurs when there is no NN model in memory for $A_i$ (\ie $M_i \cap M^\ast = \varnothing$).

\begin{equation}
\label{eq:cold_start}
C_i(M^\ast,t) =
\begin{cases}
    r_i(t)  & M_i \cap M^\ast = \varnothing \\
    0 & otherwise
\end{cases} 
\end{equation}

Assume that utilizing $m^\ast_i \in M_i$ results in an inference accuracy that we denote it as $\chi^\ast_i$. Then, based on Equation~(\ref{eq:bi_objective}), for $n$ multi-tenant DL applications, we can formally state the objective function as minimizing the total number of cold-start inferences, while maximizing the accuracy of the inferences. In this case, the total memory size available for the NN models (denoted $S$) serves as the constraint. 

% let $r_i(t)$ a Boolean function that represents the request for $A_i$ at time $t$ with value 1. Also, let $M^\ast = \{ m^(i)_j | m^(i)_j \in M_i \ \text{and} \ 1\le j \le n\}$

% Also, for NN model  $m_k \in M_i$, let $s_k$ denote its size, and $\mu_k$ denote its inference accuracy. In each request for an application $A_i$, $r_i(t) = 1$, we would like to have a NN model $m_k \in M_i$ with the best accuracy loaded in the memory to provide a warm-start inference with satisfying accuracy. 

% For application $A_i \in A$ with $M_i$ as its model zoo, let $c_i(t)$ a Boolean function that represents the cold-start inference of application $i$ at time $t$ with value 1. Also, for NN model $m_i\in M_i$ that is loaded into the memory, let $s_i(t)$ denote its size, and $\mu_i(t)$ denote its inference accuracy. We have $c_i(t) = 1$, if and only if $s_i(t)=0$. Then, as noted in Equation~(\ref{eq:1}), for $n$ multi-tenant DL applications, we can formally state the objective function as minimizing the total number of cold-start inferences, while maximizing the accuracy of the inferences. In this case, the total memory size available for the NN models (denoted $S$) serves as the constraint.
\begin{equation}
\label{eq:bi_objective}
\begin{gathered}
    \min \biggl( \int_{t}^{\infty} \sum\limits^{n}_{i=1}   C_i(M^\ast, t) \ dt \biggr) \ , \quad
    \max \biggl( \int_{t}^{\infty} \sum_{i=1}^{n} \chi^\ast_i(t)  \ dt  \biggr)\\
    \textrm{subject to:} \quad \quad \quad
    \forall t ,\ \ \sum\limits_{i=1}^{n} s^\ast_i \le S \quad \quad \quad 
     % \forall{A_i} \in A \ \exists m_i \in M_i \ \  \textrm{s.t.} \ \ s_i(t) = size(m_i) \\ 
\end{gathered}
\end{equation}

% \begin{equation}
% \label{eq:bi_objective}
% \begin{gathered}
%     \min \int_{t}^{\infty} \sum\limits^{n}_{i=1}   c_i(t) \ dt \ , \quad
%     \max \int_{t}^{\infty} \sum_{i=1}^{n} \mu_i(t)  \ dt \\
%     \textrm{subject to:} \\ 
%     \forall t ,\ \ \sum\limits_{i=1}^{n} s_i(t) \le S \\ 
%      \forall{A_i} \in A \ \exists m_i \in M_i \ \  \textrm{s.t.} \ \ s_i(t) = size(m_i) \\   
% \end{gathered}
% \end{equation}
% the goal is to load models into the memory efficiently such that the cold-start inferences are prevented while the model accuracy is satisfactory. 
Note that optimal NN model management decisions do not have a greedy nature. That is, minimizing the number of cold-start inferences at a given time $t$ does not necessarily lead to the minimum total number of cold-starts with maximum accuracy during the entire applications' lifetime. In other words, the system may experience a cold-start at time $t$ to prevent multiple ones at a later time. That is why, the objective function of Equation~\ref{eq:bi_objective} includes integrals over $t$ to the $\infty$ to encompass the impacts of the decisions at $t$ on the future cold-starts and accuracy levels. In the objectives, the NN models of application $A_i$ are only chosen from its model zoo ($M_i$), thus, the accuracy ($\mu_i(t)$) and size functions ($s_i$) are discrete functions. It is needless to say that \emph{minimizing the number of cold-start inferences} is equivalent to \emph{maximizing the number of warm-start events}~\cite{manner2018cold}. In the rest of this study, we use these two interchangeably.

\subsection{Solution Statement and Contributions}
To stimulate multi-tenancy on the limited edge memory, we develop a framework, called \name, that takes advantage of a model zoo for each DL application and can dynamically swap the NN models of the applications. To maximize the number of warm-starts with high inference accuracy across multi-tenant DL applications, our approach is to proactively load the high-precision NN models for the applications that are expected to receive inference requests, while loading low-precision models for the others. We utilize the recent memory usage information to predict the memory availability for the next executions while not interrupting other active applications. We develop model management heuristic policies that make use of the expected memory availability and the usage pattern of multi-tenant DL applications to choose a suitable NN model for the requester application right before the inference operation, thereby, both the latency and inference accuracy of the application are fulfilled. 

In summary, the contributions of this work are as follows:
\begin{itemize}
    \item We develop an NN model management framework for multi-tenant DL applications on the edge server, called \name, that efficiently utilizes the memory such that the multi-tenancy degree and number of warm-start inferences are maximized without any major compromise on the inference accuracy. \name dynamically loads the high-precision NN model for the requester application, while loading low-precision ones for others.
    %\item We develop a simulator of \name~framework and make it open source to support future research. %for latency-sensitive concurrent execution of smart applications with maximum accuracy on a resource constrained edge device.
    \item  We develop iWS-BFE policy along with three other baseline heuristic policies within \name to choose the suitable model for the application performing inference, and to decide how to allocate memory for it. %The heuristicfunctions based on the Bayesian theory to choose to load the appropriate NN model based on the likelihood of incoming inference requests from the IoT device.
   % functions based on the Bayesian theory and proactively loads the NN model that maximizes the inference accuracy likelihood of incoming inference requests from the IoT device.
    \item We evaluate and analyze the efficacy of the proposed heuristics in terms of their effectiveness and robustness against uncertainties in the inference request prediction.
\end{itemize}

The rest of this paper is organized as follows. Section \ref{sec:relwk} discusses background study and related prior works. We explain the overview of \name architecture in Section \ref{sec:propo}. Next, we discuss experimental evaluation and performance analysis in Section \ref{sec:evltn}. Finally, Section \ref{sec:conclsn} concludes the paper and provides a few avenues for the future studies.

%% file: Sources/sec-rw.tex
\section{Background and Related Work}
\label{sec:relwk}

\subsection{Edge AI}

\subsubsection{The Scope for Edge Intelligence}
Numerous research have been undertaken to explore the applications, scopes, and benefits of edge-based AI for the seamless execution of latency-sensitive smart applications~\cite{chen2019deep,murshed2021machine,deng2020edge,zhou2019edge}.
Murshed~\etal discussed different DNN-based practical applications such as video analytics and image
recognition for enabling edge AI~\cite{murshed2021machine}. 
Zhou~\etal  surveyed on various training and inference techniques for NN models on edge devices~\cite{zhou2019edge}. Chen and Ran discussed different techniques that can help to accelerate the DL training and inference on the edge-based systems~\cite{chen2019deep}. Han~\etal explored the ways to accelerate the training convergence for the edge-based architectures~\cite{deng2020edge}.
%As multiple computing tiers are involved in edge-based systems, several researchers have focused on the communication aspects and multi-party (a.k.a. federated) training techniques.  
%Lim~\etal provided a comprehensive analysis of the efficacy of federated learning in deploying DL-based smart applications on edge systems~\cite{lim2020federated}.
%Park~\etal focused on realizing edge AI over wireless channels. 
Wang~\etal surveyed the development of DL applications on edge from the latency and bandwidth perspectives~\cite{wang2020convergence}.
Zhou \etal~\cite{zhou2019edge} claimed that although higher edge intelligence reduces data offloading and improves the privacy, the latency and energy consumption overhead can increase.

%[ABOUT ZHOU]categorized edge intelligence in two six levels that are, namely cloud-edge co-inference and cloud training, in-edge co-inference and cloud training, on-device inference and cloud training, cloud-edge co-training and inference, all in edge training and inference, and all in on-device training and inference. They reported that although higher edge intelligence reduces data offloading and improves the privacy, the latency and energy consumption overhead can increase.

\subsubsection{Multi-tenant Execution on Edge}
Prior studies investigated AI multi-tenancy on the edge servers. Mao~\etal proposed a mobile computing framework, MoDNN, to execute DL applications simultaneously on resource-constrained devices~\cite{mao2017modnn}. MoDNN can
partition pre-trained DNN models across several mobile devices to accelerate tensor processing with reduced device-level computing cost and memory usage while achieving $2.17\times$---$4.28\times$ speedup.
%Upon deployment on four LG Nexus 5 mobile devices, MoDNN can speedup the DNN processing by $2.17\times$---$4.28\times$.

Multi-tenant execution across edge servers can lead to undesirable latency in application execution.  Ko~\etal proposed DisCo, a multi-tenant DL application execution offloading framework that enables execution of both the compute- and data-intensive parts of applications either on the device or on the edge \cite{ko2017disco}. Hadidi~\etal discussed that complex DNN models are sensitive to data loss as they depend more on the nuances in the data~\cite{hadidi2018distributed}. They mentioned losing one layer of the Inception V3 model can deteriorate the accuracy by more than $50\%$. 
They utilized distributed DNN models on IoT systems to reduce the processing and the memory footprints. %This technique reduces the impact of data loss and mean response time by not waiting for the data incoming from any straggler IoT device. 
% Subsequently, low-latency on demand executions were achievable with higher degree of inference accuracy.

The aforementioned research works addressed the problem of accelerating multi-tenant applications without considering the memory constraint of the edge servers. The only exception, to the best of our knowledge is \cite{xie2017towards}, in which the authors explored the executing the obstacle detection application in an autonomous vehicle with ultra low-latency constraint upon compromising with other executing applications. They proposed a reinforcement learning-based technique to scavenge memory from a non-priority application, hence, executing the obstacle detection application immediately and avoid accidents. Although their technique is effective to serve the latency-sensitive task, multi-tenant executions is out of their scope~\cite{xie2017towards}. 
In contrast to these works, we investigate the problem of memory management to increase the degree of multi-tenancy and the number of warm-start inferences, thereby, improving the practical usability of IoT-based systems.

\subsection{DNN Model Compression}
Model compression techniques allow for running a model on different resource-constrained devices. There are mainly two techniques to reduce the complexity of a given DNN model: making use of a fewer bit widths (a.k.a. \emph{quantization}) and using fewer weights (a.k.a. \emph{pruning}). These techniques have been considered individually and together to serve the purpose of model compression.  

\noindent{\textbf{Quantization.}}
Quantization reduces the computational resource demand at the expense of a diminutive loss in accuracy. By default, the model weights are float32 type variables which means 4 bytes are associated with each model weight with a significant amount of memory requirements. Model weights can be reduced from 32 bits to 8 bits (or even shorter~\cite{gholami2021survey}) to accelerate inference operation.
%It is possible to reduce the model weights from 32 bits to 8 bits (or even shorter \cite{gholami2021survey}) and gain $4\times$ reduction in the model size that has to be loaded into the memory for the inference operation.

\noindent{\textbf{Pruning.}}
The pruning technique is applied %during the inference
to reduce the memory consumption of the model to accelerate the inference operations. An effective pruning technique removes redundant connections and/or reduces the width of a layer while ensuring a slight impact on the inference accuracy. Therefore, the pruned models are retrained to compensates the loss in accuracy. 
Failure of selecting proper pruning candidates affects inference tasks and make the pruned model futile. Some studies have also been conducted on the selection of appropriate pruning candidates.

%Yang~\etal prioritized pruning candidates based on their energy consumption~\cite{yang2017designing}. They proposed 
%energy evaluation technique to estimate the DNN energy that accounts for the data movement from different levels of the memory hierarchy. However, this technique could potentially prune connections that affect the inference performance of the DNN model. 

For compatibility with the IoT devices, Yao~\etal proposed DeepIoT~\cite{yao2017deepiot}, a reinforcement learning pruning technique for DNN models in the IoT devices. However, during pruning the model parameters, %(\ie weights), 
they only considered the execution time speed-up, hence, the technique inevitably exhibits inferior inference accuracy performance.
As noted above, aggressive pruning often
substantially degrades the inference accuracy. Training and
inference with high pruning with negligible impact on the accuracy is still an open research problem~\cite{gholami2021survey}.

\noindent{\textbf{Warm-Start vs Cold-Start Inference.}}
Provided the increasing complexity of DNN models, loading even compressed models to the edge memory is a burden. The problem is further complicated in scenarios where the edge server has to continuous maintain multiple applications in its memory (\ie multi-tenancy) which is cost-prohibitive.
%In such scenarios, permanently keeping a model in-memory to conduct warm-start inference is cost-prohibitive. 

Nonetheless, cold-start inferences should be avoided as they bring about a remarkable inference latency (see \textit{loading time} in Table \ref{tab:model_des} for more details). Some research works have been accomplished to avoid cold-start inferences. For instance, to support latency-sensitive applications, in~\cite{ma2019moving}, the authors proposed cold-start of a DNN model in the background while the user is browsing a specific web page. By utilizing system resources, their technique tracks the user's browsing activity and loads the task-specific model in parallel during browsing activity to avoid the cold-start.

%% file: Sources/sec-prop.tex
\section{Allocating Multi-tenant Deep Learning Applications on the Edge Systems}
\label{sec:propo}
%\subsection{Performance Evaluation on a Secure Cloud}
% 

\subsection{Architectural Overview \& System Design of \name}
\begin{figure}
  \centering
    \includegraphics[width=0.47\textwidth]{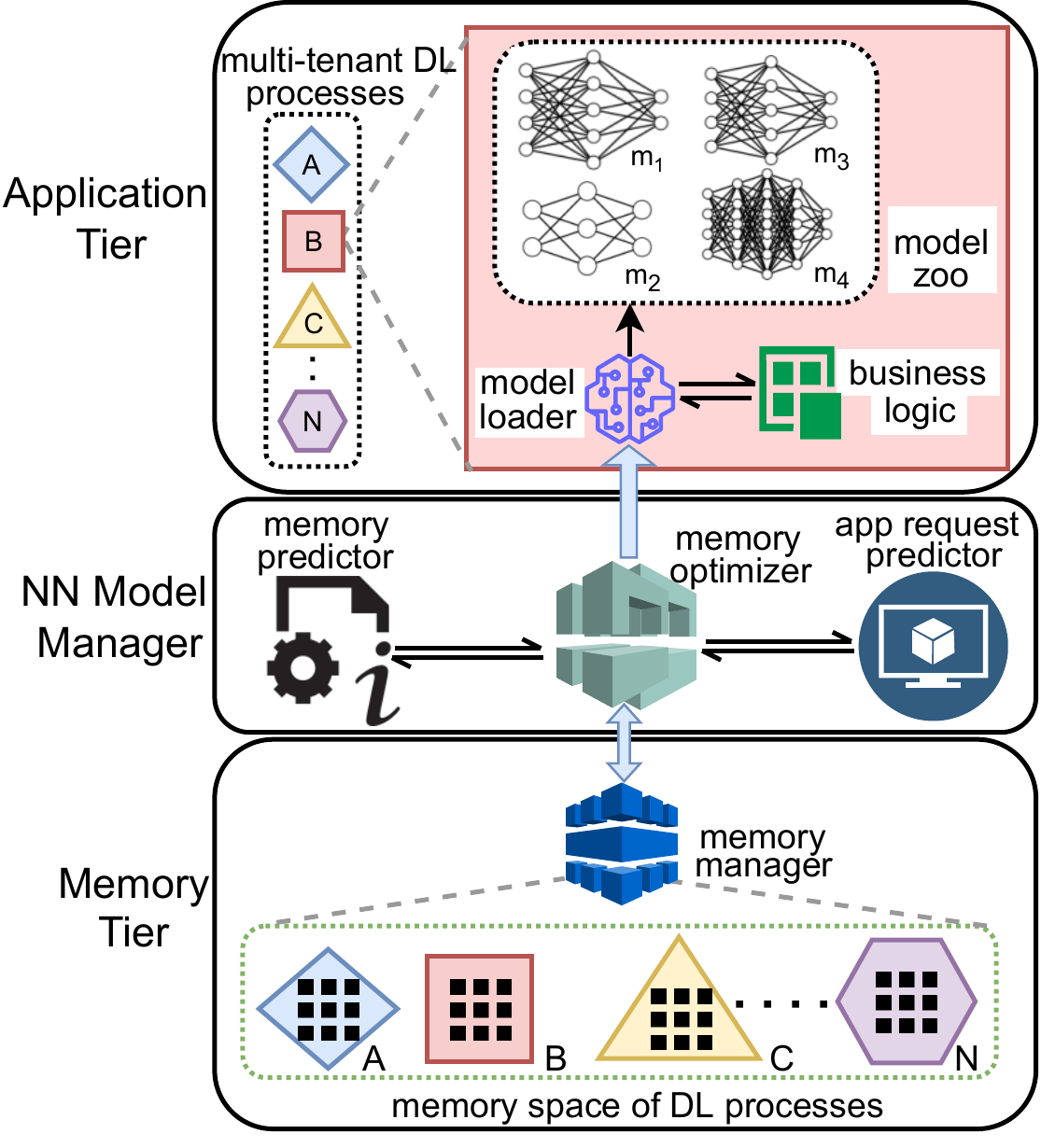}
  %\end{center}
  \caption{\small{Architectural overview of the \name~framework with three tiers: Application, NN Model Manager, and Memory.}}\label{fig:archi}
  \vspace{-5 mm}
\end{figure}

% Figure~\ref{fig:archi} represents a high-level architectural overview of \name~framework for simultaneous executions of DL applications. 
Figure~\ref{fig:archi} illustrates the architectural overview of \name that facilitates multi-tenancy of DL applications on a resource-limited edge system via enabling the applications to only swap their NN models, instead of the entire application. The framework consists of three tiers: (i) Application tier, (ii) NN model manager, and (iii) Memory tier.

    %\begin{wrapfigure}{r}{0.55\textwidth}
%In the figure, the edge system is divided into three parts: \emph{Application tier; NN model manager; and memory tier}. 

\noindent\textbf{Application Tier.} The incoming multi-modal inputs from the connected IoT devices trigger execution of multi-tenant DL applications in the application tier. The \emph{model zoo} for each DL application acts as a repository that contains NN models with different compression levels (sizes) and inference accuracy (a.k.a. various \emph{precision levels}). The \emph{model loader} is responsible for loading the chosen NN model from the model zoo into the edge memory. 

% The application tier contains a set of simultaneously executable multiple multi-tenant DL application processes incoming from connected IoT devices. Alongside the business logic and data, each application also contains \emph{model zoo} and \emph{model loader}. The \emph{model zoo} acts as a repository containing different sized models and the \emph{model loader} loads a model on the memory from the model zoo.

\noindent\textbf{NN Model Manager.} NN model manager comprises of three components: (i) \emph{application request predictor}, (ii) \emph{memory predictor}, and (iii) \emph{memory optimizer}. ``Application request predictor'' collects historical requests to each application and trains a lightweight (edge-friendly) many-to-one vanilla recurrent neural network (RNN) time series prediction model, similar to the one in \cite{pang2020innovative}, to periodically foresee the inference request arrivals for each application. 
Upon arrival of each request, ``memory predictor'' is in charge of predicting the memory availability based on the recent memory allocations in the entire edge system. We leverage the historical memory allocation data and train another many-to-one vanilla RNN time-series prediction model to predict the available memory.

Memory optimizer interacts with the 
application ``request predictor'' and ``memory predictor'' to receive: (A) the request arrival time for different applications plus the information of their model zoo; and (B) the memory availability information. Then, the memory optimizer feeds the received information to an NN model management policy that determines the highest possible precision NN model that can be loaded to serve the inference request of a DL application with the minimum impact (in terms of the prediction accuracy or latency) on the execution of other applications. Upon facing memory shortage for an arriving inference request, the memory optimizer scavenges the memory allocated to the NN models of other applications via either loading a lower-precision model or forcing them to cold-start. After procuring adequate memory, the memory optimizer informs the ``model loader'' to load the appropriate NN model of the  requested application.

 \noindent\textbf{Memory Tier.} The tier includes the ``memory spaces'' allocated to the applications; and a ``memory manager'' that keeps track of the currently loaded models, the available memory spaces, and the current status of the applications. The memory manager communicates these information to the NN Model Manager to efficiently allocates them to the arriving requests.

\subsection{Heuristics to Manage Models of Multi-tenant Applications}
\subsubsection{Overview}
Recall that the aim of NN model management policy is to minimize the number of cold-start inferences and maximize the inference accuracy for multi-tenant DL applications on the edge servers. To that end, the memory optimizer strives to maximize the time to retain the loaded models in the edge memory. However, due to limitations in the available memory space, it is not possible to retain the highest precision NN model of all applications in the memory. To resolve this memory contention, the NN models of the applications that are %maintaining models of recently executed applications or 
unlikely to be requested in the near future should be assigned a lower priority to remain in the memory. Furthermore, \name makes it possible to dynamically load NN models for the applications. This means that, upon predicting time $t$ as the inference request time for a given DL application, \name can be instructed to load the high-precision NN model of that application immediately before performing the inference. Similarly, in the face of a memory shortage, for the application(s) that are unlikely to be requested at time $t$, \name can be instructed to unload their NN models or, more interestingly, replace them with a lower precision one. 

\begin{figure}
    \centering
    \includegraphics[width=.4\textwidth]{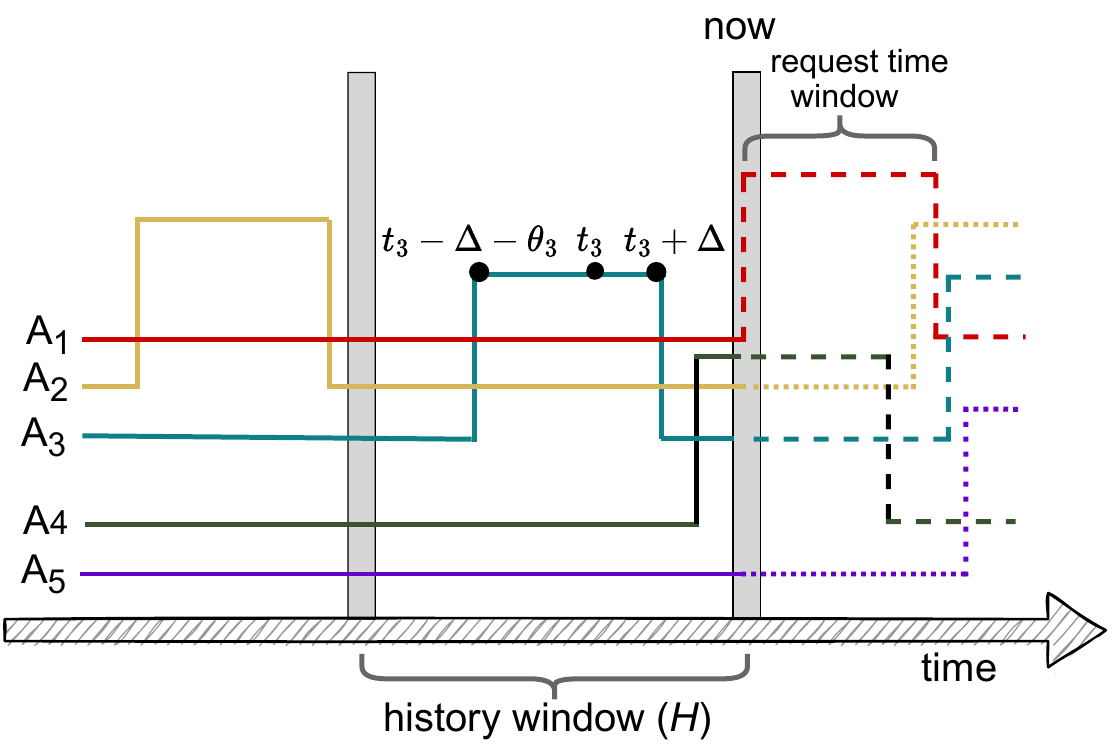}
    \caption{A sample scenario of inference requests for five multi-tenant applications, namely $A_1$ to $A_5$. Each pulse represents the time window within which an inference request is expected. Solid lines expresses the event that has already happened and dashed lines after ``now'' are the request predictions.}
\label{fig:sim}
\vspace{-5 mm}
\end{figure}

However, we know that the request arrivals are inherently uncertain \cite{felare2022} and no prediction model can precisely capture the exact request time for an application. To capture the uncertainty, we consider a request time window, denoted as $\Delta$, around each predicted request time. %obtained from the \emph{application request predictor}.
The value of $\Delta$ is obtained from profiling past request predictions and calculating the mean difference of actual arrival time and the predicted ones across all applications.  
In addition, there is a time overhead, denoted as $\theta_i$, to load the chosen NN model of an application $A_i$ into the memory. In sum, to prevent a cold-start for $A_i$ that is predicted to perform inference at time $t$, as shown in Figure~\ref{fig:sim}, the NN model  has to be loaded at time $(t_i-\Delta-\theta_i)$ and kept in memory until $(t_i+\Delta)$.    
Furthermore, there is uncertainty in predictions of ``no request'' for an application at a given time. That is, at time $t$, there can be an inference request for an application that was predicted not to have an request at that time. To make the system robust against this type of uncertainty and to avoid cold-start inferences in these circumstances, an ideal policy should load low-precision NN models for these applications. Hence, an unpredicted inference request can be still served as a warm-start by the low-precision model and the latency constraint is maintained. 
% Furthermore, to make the system robust against uncertain predictions, we propose to keep the lightweight model in memory based on memory availability. In this way, an unpredicted application request can be served still the lightweight model will serve the application usage while maintaining latency sensitivity. Keeping the lightweight models in warm-start while the related applications are not being executed is defined as \emph{minimalists} (denoted $A^\prime$).

In this work, the set of applications whose NN models are retained in memory outside of their predicted request time window are called the \emph{minimalist} set, and denoted as $A^\prime$. Similarly, the set of applications that are in their request time window and we load a high-precision model for them are called \emph{maximalist}, and denoted as $A^\ast$. 
To resolve the memory contention, the policy can be based on scavenging memory from the minimalist applications to procure the required memory space for the maximalist ones. That is, in the event that application $A_i$ is predicted to have an inference at time $t_i$, it becomes a member of $A^\ast$ set at time $t_i-\Delta-\theta_i$, and then becomes a member of $A^\prime$ set after $t_i+\Delta$; thus, its model can be evicted from the memory in the event the memory space is needed for another maximalist application. The NN model eviction is only permitted from $A^\prime$ set and we aim at retaining a low-precision model for the applications in this set. However, due to high inference demand, $A^\prime$ have to unload their models (\ie switch to cold-start) to free space for the model of the applications that are in the maximalist set. In an extreme situation, if $A^\prime$ is empty, or the scavenged memory from $A^\prime$ cannot procure sufficient space to load the suitable model for application $A_i$, the next (smaller) model for $A_i$ is considered, and the aforementioned steps are repeated. Ultimately, if the scavenged memory space is inadequate for the lowest precision model of $A_i$, an inference failure occurs.

The memory contention problem can be reduced to the classic binary Knapsack optimization problem \cite{chen2021fepim} where from a collection of items, each one with a weight and a value, we need to select items such that the total value is maximized, while the total weight is bounded to a limit. This problem is known to be NP-Complete,% problem as getting optimal solution within polynomial time is non-determined (unknown) whereas the solution is verifiable within polynomial time. 
hence, we can rely on the heuristic-based solutions for it \cite{christensenheuristic}. In the next part, we discuss four NN model management (a.k.a. \emph{NN model eviction}) policies to manage the memory for multi-tenant DL applications such that the number of warm-start inferences is maximized without any major impact on the inference accuracy.

\subsubsection{Policy 1: Largest-First Model Eviction (LFE)}
In this policy, to allocate memory for the NN model of a maximalist process, we first evict NN models from set ($A^\prime$ that occupy the highest memory space, until there is enough space to allocate the high-precision NN model of $A^\ast$. For that purpose, members of $A^\prime$ are sorted based on the size of their currently loaded NN model in the descending order. In the event that evicting all the NN models of $A_\prime$ does not free enough memory space to allocate the NN model of the request, a lower precision NN model (smaller in size) is tried for allocation. This procedure continues until a model from the model zoo can be allocated in the memory; otherwise, the edge system is not able to serve that request at that time. %Note that an application whose model is evicted directly goes to cold-start for the unpredicted future application request.

\subsubsection{Policy 2: Best-Fit Model Eviction (BFE)}
The limitation of LFE is to evict the largest NN models of the minimalist applications, irrespective of the exact memory requirement. This means that adopting LFE  can free more memory space than the actual requirement. To tackle the issue, we implement the BFE policy where applications in the minimalist set are sorted based on the difference between their model sizes and the actual memory requirement. Then, the NN model with a minimum difference is chosen for eviction. 
% \begin{equation}
%     sort {A^\prime} based on {M_j^{A_i} \forall {A_1.... A_n} - A_i} 
% \end{equation}
The memory requirement for a maximalist application is first calculated based on its highest precision (largest) NN model to gain the highest inference accuracy. However, in the event that evicting the NN models of all the minimalist applications do not free enough memory space to allocate the desired NN model, BFE iteratively selects the next high-precision model from the model zoo of the requested application. 

\subsubsection{Policy 3: Warm-Start-aware Best-Fit Model Eviction (WS-BFE)} 
Let $A_i\in A^\ast$ an application that is currently in the maximalist set, and $A_j\in A^\prime$ an application that is currently in the minimalist set. It is technically possible that the predicted request time window of $A_i$ overlaps with the one for $A_j$. In this case, LFE and BFE policies potentially choose to evict the NN model of $A_j$ in favor of the $A_i$ model. This is because both of these policies are backward-looking and ignore the fact that $A_j$ can be requested soon after evicting its NN model. Such an eviction decision increases the likelihood of a cold-start inference and to avoid that, we develop WS-BFE that assigns the lowest eviction priority to those applications in $A^\prime$ that have overlapping time window with $A_i$. 

In our early experiments, we realized that another reason for cold-start inferences is due to uncertain nature of request arrivals. That is, a minimalist application is unexpectedly requested. To minimize the likelihood of cold-start inference in these circumstances, we implement WS-BFE to \emph{replace} the evicted NN model with the lowest-precision (\ie smallest) NN model of that application. As such, in the event of an unpredicted request the minimalist applications, there is a low-precision model available to carry out a warm-start inference. 
%Ali, Please provide a formulation about Ws-BFE.

\subsubsection{Policy 4: Intelligent Warm-Start-aware Best-Fit Eviction (iWS-BFE)} To make WS-BFE robust against uncertainties in the application request time prediction, we enhance it by applying the Bayesian theory and proposing a new policy, called iWS-BFE. This policy is inspired from the widely-adopted LRU-K cache management policy~\cite{o1993lru} that considers \emph{the least recently used (\ie requested) applications are not likely to be requested in the near future}. %In~\cite{o1993lru}, the authors provided three different experiments namely two pool, Zipfian random access, and OLTP trace with different settings  to prove the effectiveness of LRU.   
Similarly, iWS-BFE only considers members of $A^\prime$ as eviction candidates, denoted by $E^\prime$, that are not recently requested.
% Similarly, iWS-BFE prioritizes the applications of set $A^\prime$ based on their likelihood of invocation in the near future.
%Subsequently, we assume the historical data of all executions are collectable to compute prior probabilities of all executions.% (denoted $P(A_1), P(A_2), P(A_3),..., P(A_n)$ ).   
Figure~\ref{fig:sim}, shows a scenario of predicted request times for $A_1$---$A_5$. To procure memory for $A_1$, we have $A^\prime=\{A_2, A_3, A_5\}$. Because $A_3$ was requested during the ``history window'' ($H$), it is likely to be requested in the near future. Hence, iWS-BFE, chooses $E^\prime=\{A_2, A_5\}$ for eviction. The value of $H$ is determined based on the mean request inter-arrival time of all applications.  

In addition to considering LRU, iWS-BFE also makes use of the request prediction, provided by \name. That is, it considers the most appropriate application for eviction as the one that has not been recently requested, and is predicted to be requested the latest in future. However, the request time predictions are uncertain, and the system can receive an unexpected request from members of $E^\prime$ in the current request window. To make iWS-BFE robust against such uncertainty, we calculate the probability of an unexpected request. For application $A_j\in E^\prime$, let $r_j$ denote an unexpected request. Then, the probability of $r_j$ occurring during the current request window (\ie [$t, t+\Delta$]) is defined as $P(r_j | A_i \in A^\ast)$. The application that is likely to be requested unexpectedly is not an optimal choice for eviction. Therefore, in Equation~\ref{eq:scoreiws}, to calculate the fitness score of $A_j$ for eviction (denoted $Score(A_j)$), we consider $1 - P(r_j | A_i \in A^\ast)$. To take the predicted request time of $A_j$ into consideration, we calculate the distance between its predicted request time and the current time (\ie $t_j-t_i$). To confine the value between [0,1], we normalize the distance based on the latest predicted distance across all $k$ applications. %Note that, in Equation~\ref{eq:scoreiws}, a higher eviction score implies a higher priority for the application to be evicted. 
% In iWS-BFE, to allocate memory for $A_i$, we need to compute the fitness of eviction candidates with respect to the current maximalist applications $A^\ast$. We utilize Bayesian theorem to prioritize eviction candidate(s), so that we can evict the application(s) with the minimal impact on serving the next inference requests. We calculate probability of an eviction candidate  

%Ali please check 
% To measure the significance of to-be evicted candidates, we  calculate the probability of any minimalist application $A^\prime_j$ %(denoted $P(A^\prime_j|A^\ast)$ ) 
% being kept in-memory given that $A_i$ has been added into the maximal state and subtract it from 1 (denoted $1-P(A^\prime_j|A_i)$).  
% To give more precedence to the to-be evicted application that is predicted to be requested later, we multiply the %obtained probability of $A^\prime_j$ 
% subtracted result with
% the normalized predicted request time of $A^\prime_j$. This provides us with the fitness scoring metric, denoted $F_{A^\prime_{j}}$, that is shown in Equation~\ref{eq:scoreiws}.

%probability of being kept.. CHANGE it!!!!
\vspace{-3 mm}
\begin{equation}
\label{eq:scoreiws}
\begin{gathered}
Score(A_j) = \frac{t_j-t_i}{\displaystyle \max_{k\in E^\prime} (t_k - t_i)} \cdotp 
\bigl[ 1 - P(r_j | A_i \in A^\ast)\bigr] %\\  t\in [t_i, t_i+\Delta] 
\end{gathered}
 \end{equation}
% F(A^\prime_{j}) =  \bigg(1- P(A^\prime_j|A_i)\bigg)  \times \frac{t_j}{\displaystyle\max_{\forall j} (t_{j})} 

The pseudo-code of the iWS-BFE policy is provided in Algorithm~\ref{alg:memory_man}. It begins with an initial set of eviction candidates, called $\tau \subseteq A^\prime$, that is formed based on the applications that were not requested during the history window ($H$). From $\tau$, in Step $3$, a list of eviction candidates (denoted $E$) whose elements do not overlap with the request window of active application ($A_i$) is derived. Next, in Step 4, we use Equation~\ref{eq:scoreiws} to calculate the fitness score for each $E_k \in E$ and then, build a max-heap tree of $E$ based on the fitness scores (Step 5). In Steps 6---10, the policy iteratively retrieves the application with the highest fitness score (\ie the max-heap root, denoted $w$) and foresees the amount of memory that can be scavenged upon replacing its loaded model with the lowest-precision one. Once the policy finds enough memory to be scavenged such that the NN model of $A_i$ (denoted $m_i$) can be loaded, in Step 13, it enacts all the NN model replacement decisions and then loads $m_i$ in Step 14. In the event that the scavenged memory is insufficient, the policy switches to the next NN model for $A_i$ that has a lower size and accuracy (Step 17). In the worst case that even the smallest NN model of $A_i$ cannot fit in the memory, the inference request fails (Step 17)~\cite{mokhtari2020autonomous}.
%

%During scavenging, iWS-BFE iteratively loads the simpler version of  the model for evicted $C_i$ aiming for maximum warm-start    

%In the equation, $T_{(A^\prime_j)}$ denotes the predicted next request time of  $A^\prime_j$. $P(A^\prime_j|A^\ast)$ denotes the Bayesian conditional probability of $A^\prime_j$ being evicted given that $A^\ast$ are in maximal state.

\begin{algorithm}
	\SetAlgoLined\DontPrintSemicolon
	\SetKwInOut{Input}{Input}
	\SetKwInOut{Output}{Output}
	\SetKwFunction{algo}{algo}
	\SetKwFunction{proc}{Procedure}{}{}
	\SetKwFunction{main}{\textbf{ChooseCenter}}
		
	\SetKwBlock{Function}{Function \texttt{iWS-BFE($A^\prime$, $A^\ast$, $A_i$, $H$)}}{end}
	
	\Function{
	   $\tau \gets$ Select $\forall A^\prime_j \in A^\prime$ not requested during $H$\;
	   $E \gets$  Determine $\forall A^\prime_j \in \tau$ non-overlapping with request window of $A_i$ \;
	   
	$\forall E_k\in E$ calculate fitness score using Equation~\ref{eq:scoreiws}\;
       Build max-heap tree of $E$ based on fitness score\;

        \While {$size(m_i) >$ available memory}  
        {  
        $w \gets$ Extract root of the max-heap tree \;
        \textbf{If}{ $w=\emptyset$} \textbf{then} break the loop \;
        Measure memory scavenged by replacing model of $w$ with its lowest-precision one \; 
        Add scavenged amount to \emph{available memory} \;
        }

	    \If {$size(m_i) \leq$ available memory} 
	    {
            Enact NN model replacement(s) decisions \;
	    Scavenge the leftover memory to load $m_i$ \;

	    }
	    \Else %: \;
	    {

            \textbf{If} there is no model left to check \textbf{then} the inference request fails \;
             Repeat Step 6---10 with the next (smaller) model \;
	    }
	     
	}
	\caption{Pseudo-code for iWS-BFE NN model eviction policy}
	\label{alg:memory_man}
	
\end{algorithm}%%%%%

%% file: Sources/sec-evaluation.tex
\section{Performance Evaluation}
\label{sec:evltn}

\subsection{Experimental Setup and Evaluation Metrics}
To evaluate the efficacy of \name and its NN model eviction policies, we benchmarked five different DL applications, namely face recognition, speech recognition, image classification, next sentence prediction, and text classification, and recorded their real characteristics, including the model size, and the inference accuracy (shown in Table~\ref{tab:app_model_size}).
We have developed the E2C simulator that enables modeling the IoT-based systems with different characteristics and configurations, and is available publicly for the community access through our Github page\footnote{Github page of the E2C simulator: \url{ https://github.com/hpcclab/E2C-Sim.git}}. The simulator has implemented all of the NN model eviction policies, and the user can quickly deploy and examine any one of them. 

\begin{table}[]
\centering
\resizebox{.98\linewidth}{!}{
\begin{tabular}{|c|c|c|c|c|}
\hline

\textbf{Application}                  & \textbf{NN Model}                                                                    & \textbf{\begin{tabular}[c]{@{}c@{}}Bit\\ Width\end{tabular}} & \textbf{\begin{tabular}[c]{@{}c@{}}Size\\ (MB)\end{tabular}} & \textbf{\begin{tabular}[c]{@{}c@{}}Accuracy\\ (\%)\end{tabular}} \\ \hline \hline
\multirow{3}{*}{Face recognition}     & \multirow{3}{*}{VGG-Face}                                                            & FP32                                                         & 535.1                                                        & 90.2                                                             \\ \cline{3-5} 
                                      &                                                                                      & FP16                                                         & 378.8                                                        & 82.5                                                             \\ \cline{3-5} 
                                      &                                                                                      & INT8                                                         & 144.2                                                        & 71.8                                                             \\ \thickhline
\multirow{3}{*}{Image classification} & \multirow{3}{*}{VIT-base-patch16}                                                    & FP32                                                         & 346.4                                                        & 94.5                                                             \\ \cline{3-5} 
                                      &                                                                                      & FP16                                                         & 242.2                                                        & 81.3                                                             \\ \cline{3-5} 
                                      &                                                                                      & INT8                                                         & 106.7                                                        & 72.2                                                             \\ \thickhline
\multirow{3}{*}{Speech recognition}   & \multirow{3}{*}{S2T-librisspeech}                                                    & FP32                                                         & 285.2                                                        & 89.7                                                             \\ \cline{3-5} 
                                      &                                                                                      & FP16                                                         & 228.0                                                          & 77.2                                                             \\ \cline{3-5} 
                                      &                                                                                      & INT8                                                         & 78.4                                                         & 68.0                                                               \\ \thickhline
\multirow{3}{*}{Sentence prediction}  & \multirow{3}{*}{\begin{tabular}[c]{@{}c@{}}Paraphrase-Mini\\ LM-L12-v2\end{tabular}} & FP32                                                         & 471.3                                                        & 88.2                                                             \\ \cline{3-5} 
                                      &                                                                                      & FP16                                                         & 377.6                                                        & 81.7                                                             \\ \cline{3-5} 
                                      &                                                                                      & INT8                                                         & 98.9                                                         & 76.2                                                             \\ \thickhline
\multirow{3}{*}{Text classification}  & \multirow{3}{*}{Roberta-base}                                                        & FP32                                                         & 499.0                                                          & 91.1                                                             \\ \cline{3-5} 
                                      &                                                                                      & FP16                                                         & 392.2                                                        & 82.4                                                             \\ \cline{3-5} 
                                      &                                                                                      & INT8                                                         & 132.3                                                        & 76.6                                                             \\ \thickhline
\end{tabular}
}
\caption{Application-specific models with different precision variants that are experimented.}
\label{tab:app_model_size}
\vspace{-6 mm}
\end{table}

\begin{figure*} 
%\begin{wrapfigure}{r}{0.9\textwidth}
    \centering
    \includegraphics[width=.75\textwidth]{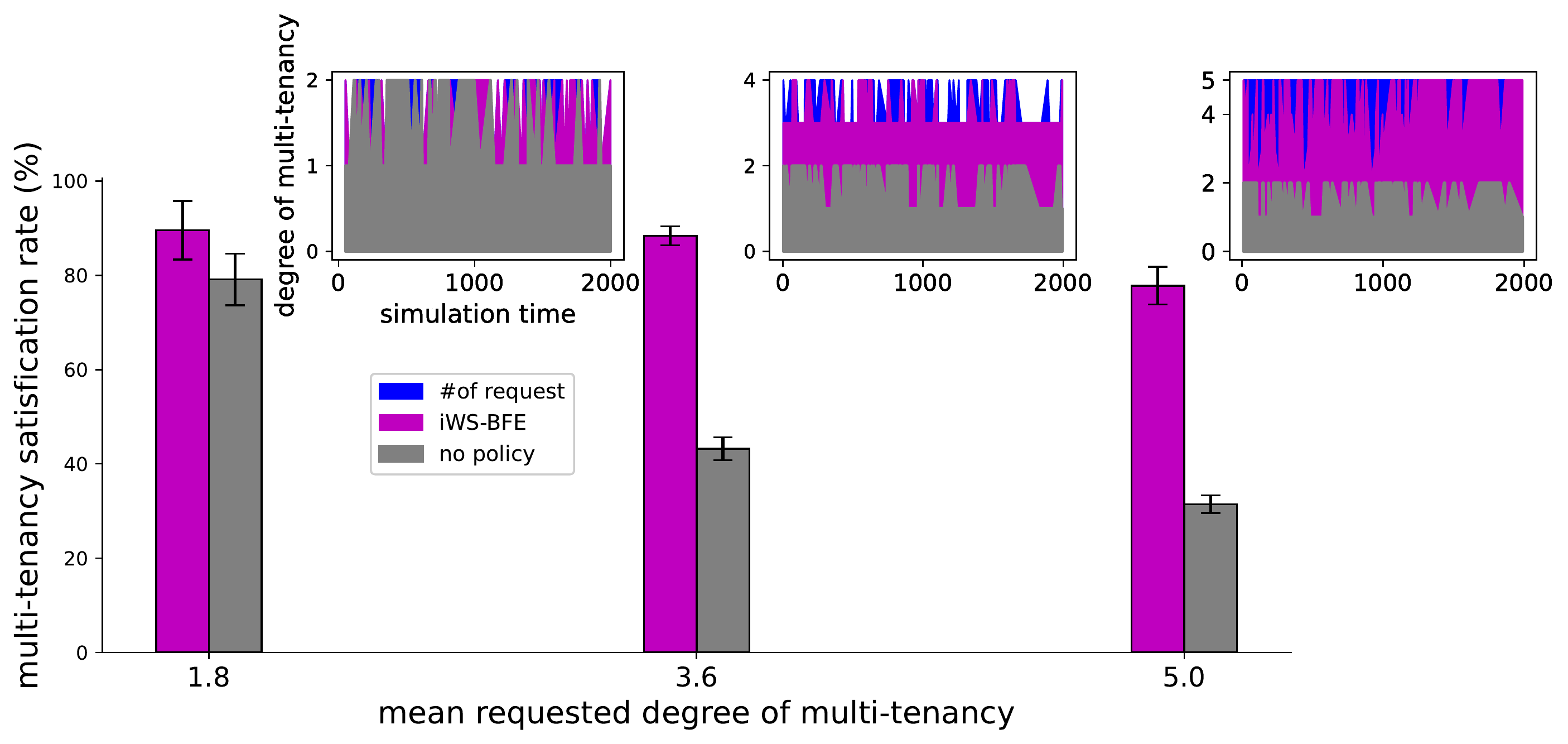}
    \caption{The impact of \name and its iWS-BFE eviction policy on satisfying the requested multi-tenancy. The large graph represents the summative analysis via increasing the mean of multi-tenancy requested in the horizontal axis, and showing the percentage of requests that were satisfied in the vertical axis. For each case, the smaller graph more granularly represents the number of concurrent requests issued and fulfilled during the simulation time.}
\label{fig:concE}
%\end{wrapfigure}
\end{figure*}
The simulator also enables us to generate workload traces that include the request arrival times for each application during the simulation time. We configure the actual workload to include an equal number of requests for the five applications, and the inter-arrival times between requests for each application are distributed exponentially within the workload.
To study the uncertainty exists in the inference request predictions, in the evaluations, we generate two sets of workloads, one includes the predicted arrival times for the multi-tenant applications, and the other one includes the actual arrival times of the applications. The distribution of request arrivals in the actual workload deviates from the distribution of requests in the predicted workload. The degree of deviation between the two is measured based on the Kullback-Leibler (KL)~\cite{van2014renyi} divergence. We explore the impact of this deviation in the experiments of next subsections.

Our evaluation metrics are: (A) The \emph{degree of multi-tenancy} under different request arrival intensity; (B) The \emph{inference latency}; (C) the \emph{inference accuracy}; and (D) The \emph{robustness} metric to measure the tolerance of different eviction policies against the uncertainty exists in the request predictions.

%\subsection{Experimental Results}
\subsection{Impact of \name on the Degree of Multi-tenancy}

This experiment is to examine the efficacy of \name~in satisfying the incoming requests to the edge server. To that end, as shown in Figure~\ref{fig:concE}, we increased the workload intensity, via the mean number of concurrent requests issued, and in each case measured the \emph{multi-tenancy satisfaction rate}, which is the percentage of warm-start inferences out of the total incoming requests during the simulation time. We examined two cases: (A) without any solution to stimulate multi-tenancy (called, no policy); and (B) with \name and its iWS-BFE policy in place. The experiment was repeated 10 times and the average rate and 95\% confidence intervals for each data point is reported.

The experiment shows that the degree of multi-tenancy achieved by adopting \name and its iWS-BFE is remarkably higher than the situation where \name is not in place. The smaller graphs show that this superiority occurs consistently during the simulation time. We also notice that the impact of employing \name is more effective for higher degrees of multi-tenancy. In particular, we can see that with the mean degree of multi-tenancy is 5, using \name and its iWS-BFE policy achieves $\approx$130\% higher satisfaction rate than no policy when mean requested degree of multi-tenancy is larger than 2. This experiment justifies the efficacy of \name and the NN model management in stimulating multi-tenancy of DL applications. 
%102\% for 3.6 least 15\%, average 83.5\%, maximum 160, 102

\subsection{Impact of the Eviction Policies on the Cold-Start Inference} The purpose of this experiment is to 
   %\begin{wrapfigure}{r}{0.55\textwidth}
   \begin{figure}
   \vspace{-10 pt}
    \centering
    \includegraphics[width=.42\textwidth]{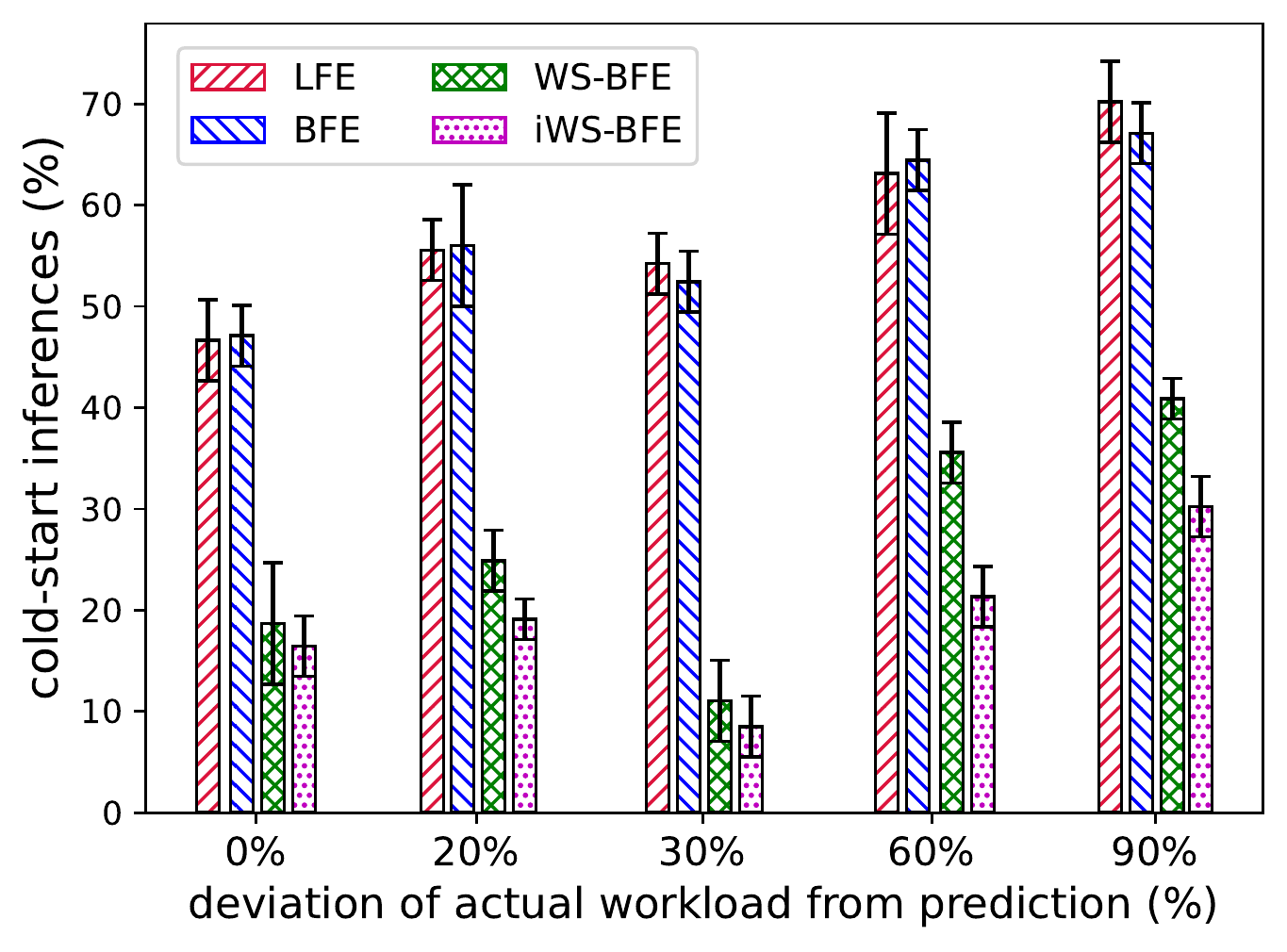}
    \caption{Measuring the percentage of cold-start inferences of multi-tenant applications resulted from %employing different NN model 
    the proposed eviction policies. The horizontal axis shows the deviation between predicted and actual inference request times.}
\label{fig:ncoldsVd}
\vspace{-5 mm}
\end{figure}
analyze the impact of different NN model eviction policies on the number of cold-start inferences. For that purpose, we measure  percentage of cold-start inferences caused by employing different eviction policies, particularly, upon varying the deviation of request prediction from the actual requests. 

The results, illustrated in  Figure~\ref{fig:ncoldsVd}, show that LFE and BFE perform poorly and cause a remarkable number of cold-start inferences, whereas, WS-BFE and iWS-BFE mitigate the cold-srart inferences by at least 65\%. This is because, in LFE and BFE, upon evicting an NN model, its corresponding application suffers from a cold-start inference in the event of an unpredicted request. In contrast, in WS-BFE and iWS-BFE, the evicted model is replaced with a low-precision one, hence, unpredicted calls to the corresponding application do not lead to cold-start inferences. It is noteworthy that, regardless of the employed policy, the percentage of cold-start inferences rises upon increasing the deviation between predicted and actual request times. Nonetheless, we see that even under 90\% deviation, iWS-BFE still substantially outperforms other policies. On average, it yields 102\% less cold-start in compare to LFE and BFE, and 40\% less than WS-BFE. 

\begin{figure}    
    \centering
    \vspace{-15 pt}
    \includegraphics[width=.42\textwidth]{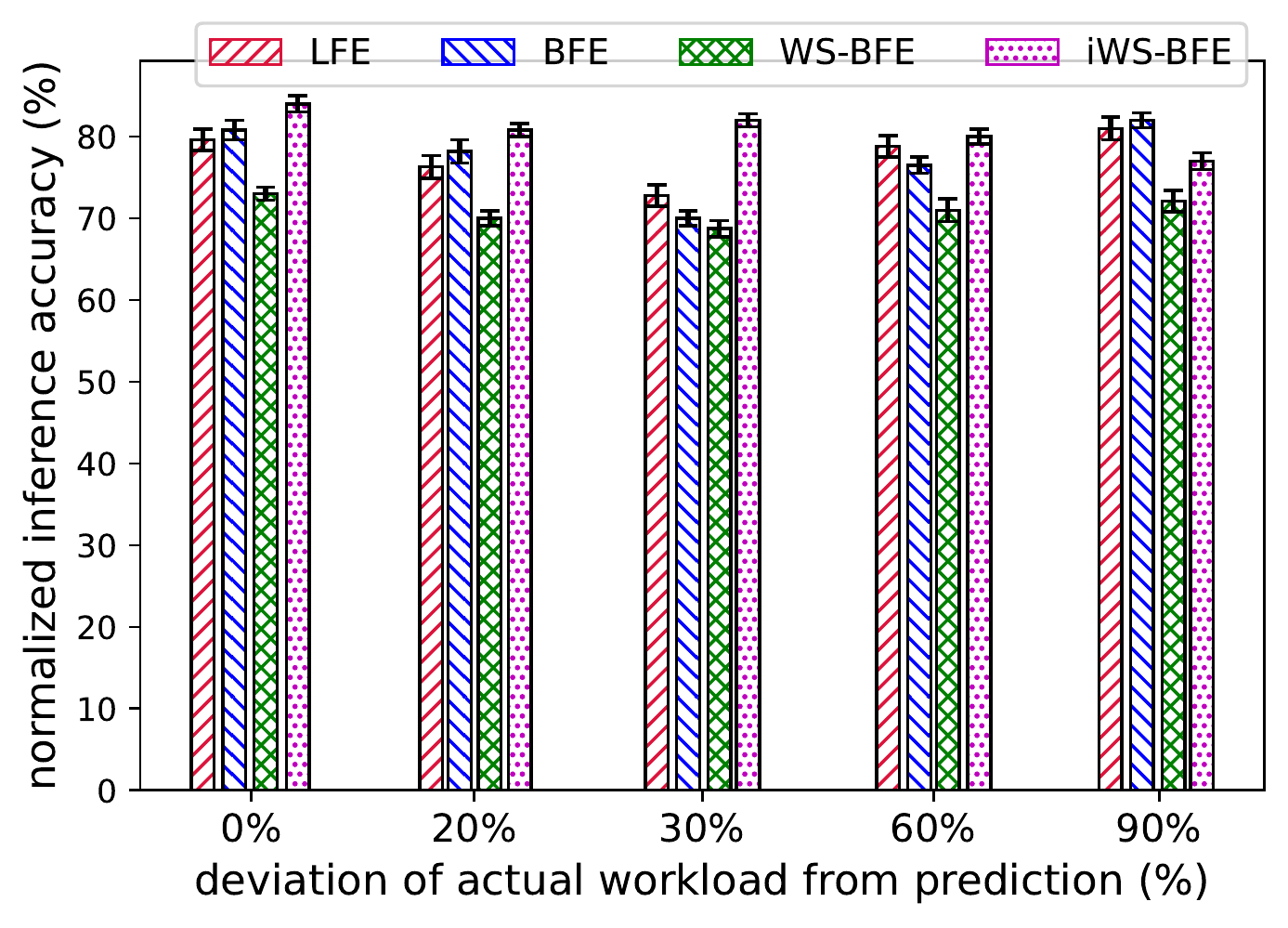}
    \caption{Measuring the normalized inference accuracy of applications resulted from employing the different eviction policies.}
\label{fig:accVd}
%\end{wrapfigure}
%\vspace{-5 mm}
\end{figure}

\subsection{Impact of the Eviction Policies on the Inference Accuracy}
In this experiment, we analyze the average inference accuracy caused by employing different model eviction policies. Because the accuracy largely varies across different applications, we perform min-max normalization on the accuracy values. %For example, INT8 variant of Roberta-base and S2Tlibrisspeech models offer 76.6\% and 68\% prediction accuracy, hence, we need to perform min-max normalization to maintain consistent accuracy across different NN models. 
Also, for the cold-start inferences, in the accuracy measurements, we  consider the accuracy provided by the NN model after it is loaded into the memory.  

%\begin{wrapfigure}{r}{0.55\textwidth}

Figure~\ref{fig:accVd} shows the normalized mean inference accuracy obtained from employing different NN model eviction policies upon changing the deviation between predicted and actual request times. According to the figure, LFE and BFE policies outperform WS-BFE. This is because, these two policies do not retain the low-precision models in the memory. Therefore, their inference requests either lead to a cold-start (that was explored in the previous experiment), or they load high-precision models that provide a high inference accuracy. Nonetheless, we observe that iWS-BFE outperforms LFE and BFE in most of the cases, except the one with 90\% deviation. The reason for the higher inference accuracy of iWS-BFE is that, it nominates cold-start candidates intelligently, based on their probability of future invocations. This results indicate the importance of the scoring (described in Equation~\ref{eq:scoreiws}) on efficiently nominating cold-start candidates. It is noteworthy that the higher inference accuracy of LFE and BFE at 90\% deviation comes with the cost of substantially higher cold-start inferences that are detrimental to the ``usability'' of the IoT-based systems.

\subsection{Bi-Objective Analysis of NN Model Eviction Policies}
\begin{figure} 
    \centering
    \includegraphics[width=.42\textwidth]{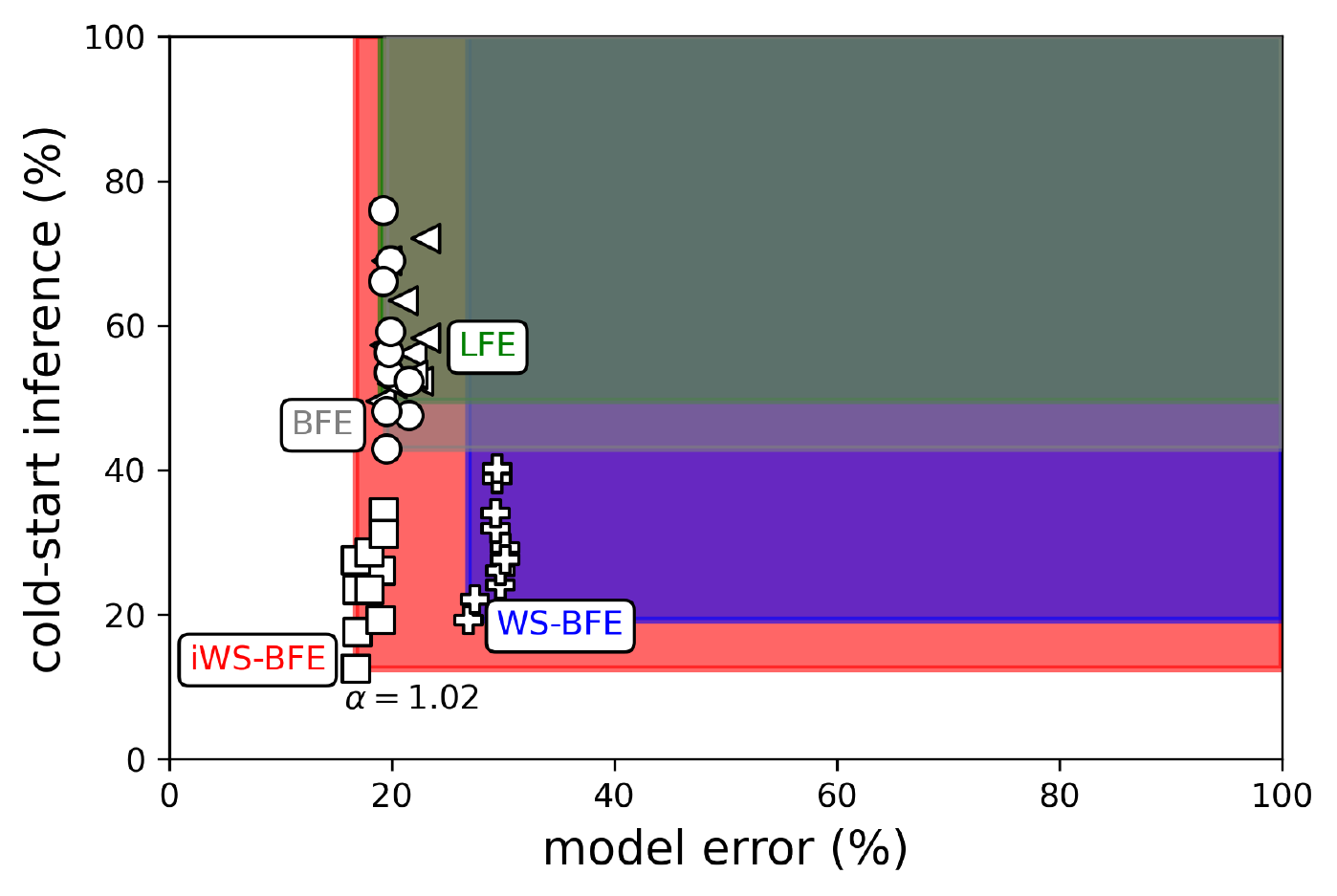}
    \caption{\small{Bi-objective analysis of the different model selection policies.}}
\label{fig:pareto}
\vspace{-1 mm}
%\end{wrapfigure}
\end{figure}

Recall that the NN model management for multi-tenant applications in a resource-limited edge system is a bi-objective optimization problem that aims at minimizing the number of cold-start inferences and maximizing the inference accuracy. However, these two are generally conflicting objectives and there is not a single optimal solution that can satisfy both objectives. Instead, there could be a range of solutions that dominate other solutions. To analyze which one of the studied policies dominate others, in Figure~\ref{fig:pareto}, we plot the percentage of cold-start inferences versus the model error (defined as 100-accuracy) for different policies  
and $\Delta$ values. Let $D$ and $\sigma$ be the mean and standard deviation of residuals of predicted versus actual request times. Then, $\Delta=D\pm\alpha\cdotp\sigma$ ranges by changing the value of $0\leq\alpha\leq2$. The deviation of actual versus predicted workload in this experiment is 30\%.

For each policy, the colored area shows the cold-start inferences and model error rate that are dominated by that policy. An ideal policy should approach the graph origin (\ie resulting in zero cold-start and zero model error). In Figure~\ref{fig:pareto}, we observe that \name dominates other policies and form the Pareto-front, particularly with $\alpha=1.02$. We can conclude that the iWS-BFE policy can significantly improve the usability of the systems via causing fewer cold-start inferences and offering a higher inference accuracy.

%This analysis recommends us to switch NN model manager to \name, particularly, at cold-start inferences greater than 15\% with model error greater than 11\%.  

\subsection{Analyzing Robustness against Uncertainties}
The goal of this experiment is to study how the eviction policies of \name~make the IoT-based system robust against the uncertainty exists between the predicted and actual application request predictor. We define the \emph{robustness metric}, shown in Equation~\ref{eq:robust}, to encompass the ratio of warm-start inferences (denoted $\varpi_{i}$) to the total number of requests (denoted $\gamma_i$), and the mean prediction accuracy ($\psi_i$) of each application $i$ throughout the simulation period. 

% formulates the way robustness is calculated. \textcolor{red}{In this equation, the sum of normalized cold-start inferences (denoted $\alpha \frac{\varpi_{c,i}}{\displaystyle\max_{\forall i} \varpi_{c,i}}$) and normalized failed inferences (denoted $\beta\frac{\varpi_{f,i}}{\displaystyle\max_{\forall i} \varpi_{f,i}}$) can reach up to 1. This is because, a failure execution of $i$ does not let any of its inference occur (either warm- or cold-start). $\alpha$ and $\beta$ are used a tuning factor and their values range from [0,1]. The sum of normalized cold-start and failed inferences are negatively impact on the performance of any policy and therefore, the sum is subtracted from $1$. The subtracted result is multiplied with $\mu_i$ as higher accuracy indicates the effectiveness of a given policy.

\begin{equation}
\vspace{ 0 mm}
\label{eq:robust}
R= \frac{1}{n} \cdotp \displaystyle\sum^n_{i=1}\bigg[{ \frac{\varpi_{i}}{\gamma_{i}}  \cdotp\psi_i } \bigg] 
\end{equation}
 \begin{figure}
    \centering
    \includegraphics[width=.45\textwidth]{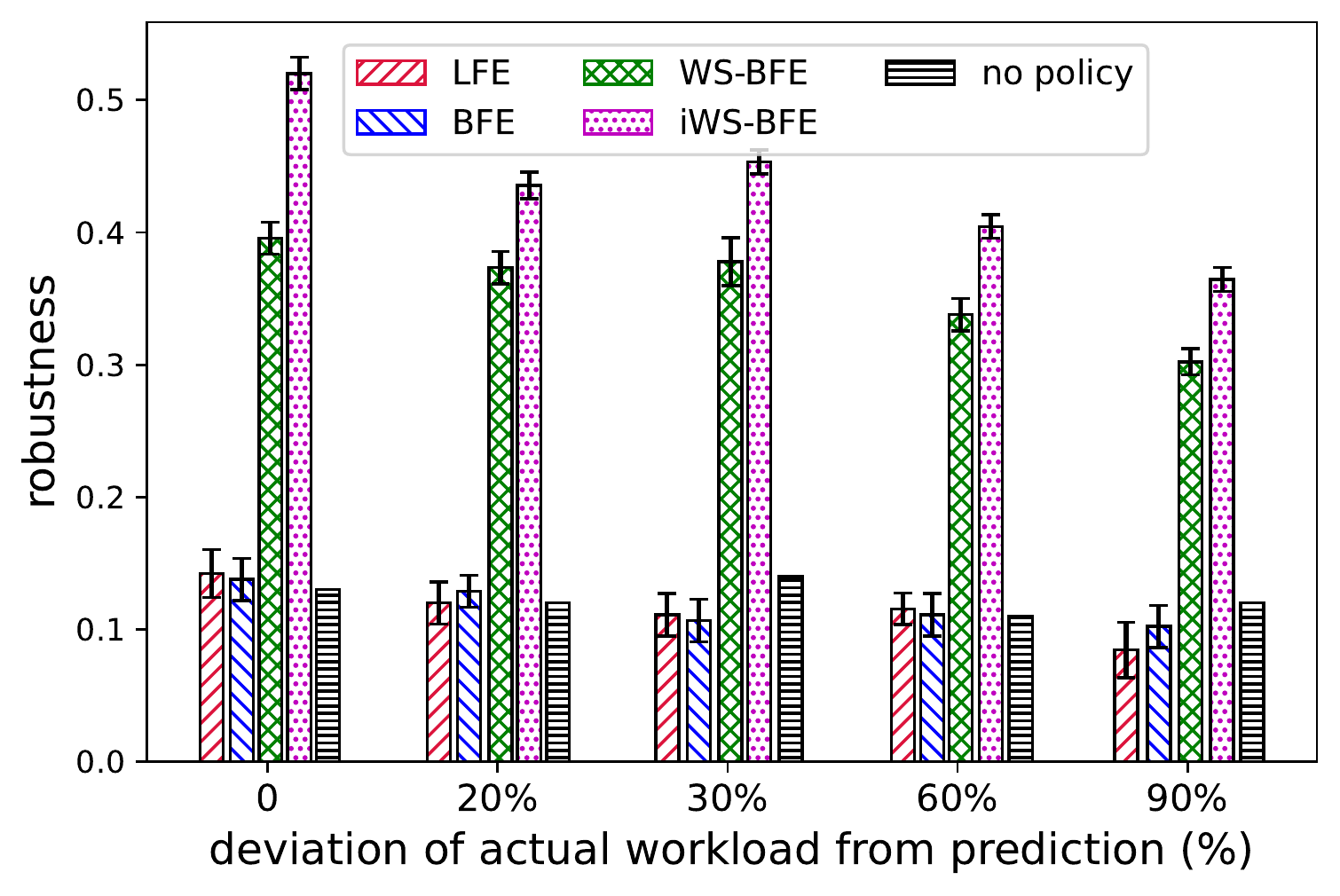}
    \caption{Robustness of the system against uncertainty in the prediction of inference requests. }
\label{fig:robust}
\vspace{- 8mm}
\end{figure}

Figure~\ref{fig:robust} represents the robustness score achieved by adopting the proposed policies and no policy (a.k.a. baseline) against uncertainties in the inference request prediction. We observe that deploying \name with any policy provides more robustness than the circumstance where \name is not in place (no policy). We also notice that the robustness value consistently drops %up to 60\% deviation and rises again for 90\% deviation. This is due to fewer inference failures of each application at 90\% deviation compared to the case of 20\%, 30\%, and 60\%.
because the rate of inference failure and cold-starts rise for higher deviations.
We observe that WS-BFE and iWS-BFE are more robust against deviation than the LFE and BFE. This is because, LFE and BFE do not replace their NN models with a lower-precision one upon eviction, which leads to cold-start inferences for the applications.

% \begin{figure} [!htbp]
% \begin{subfigure}{.8\textwidth}
% \centering
% \includegraphics[width=\linewidth]{Figures/exp/Experiment_4.pdf}
% \caption{BBC Dataset}
% \label{fig: semidyna_bbc}
% \end{subfigure}

% \hfill

% \begin{subfigure}{.8\textwidth}
% \centering
% \includegraphics[width=\linewidth]{Figures/exp/Experiment_5.pdf}
% \caption{RFC Dataset}
% \label{fig: semidyna_rfc}
% \end{subfigure}
% \caption{Clusters' coherency for different updates of the three studied datasets when SD-\name~is applied with and without re-clustering option.}
% \end{figure}

\subsection{Evaluating the Fairness of NN Model Eviction Policies}
\begin{figure} 
%\begin{wrapfigure}{r}{0.9\textwidth}
    \centering
    \includegraphics[width=.44\textwidth]{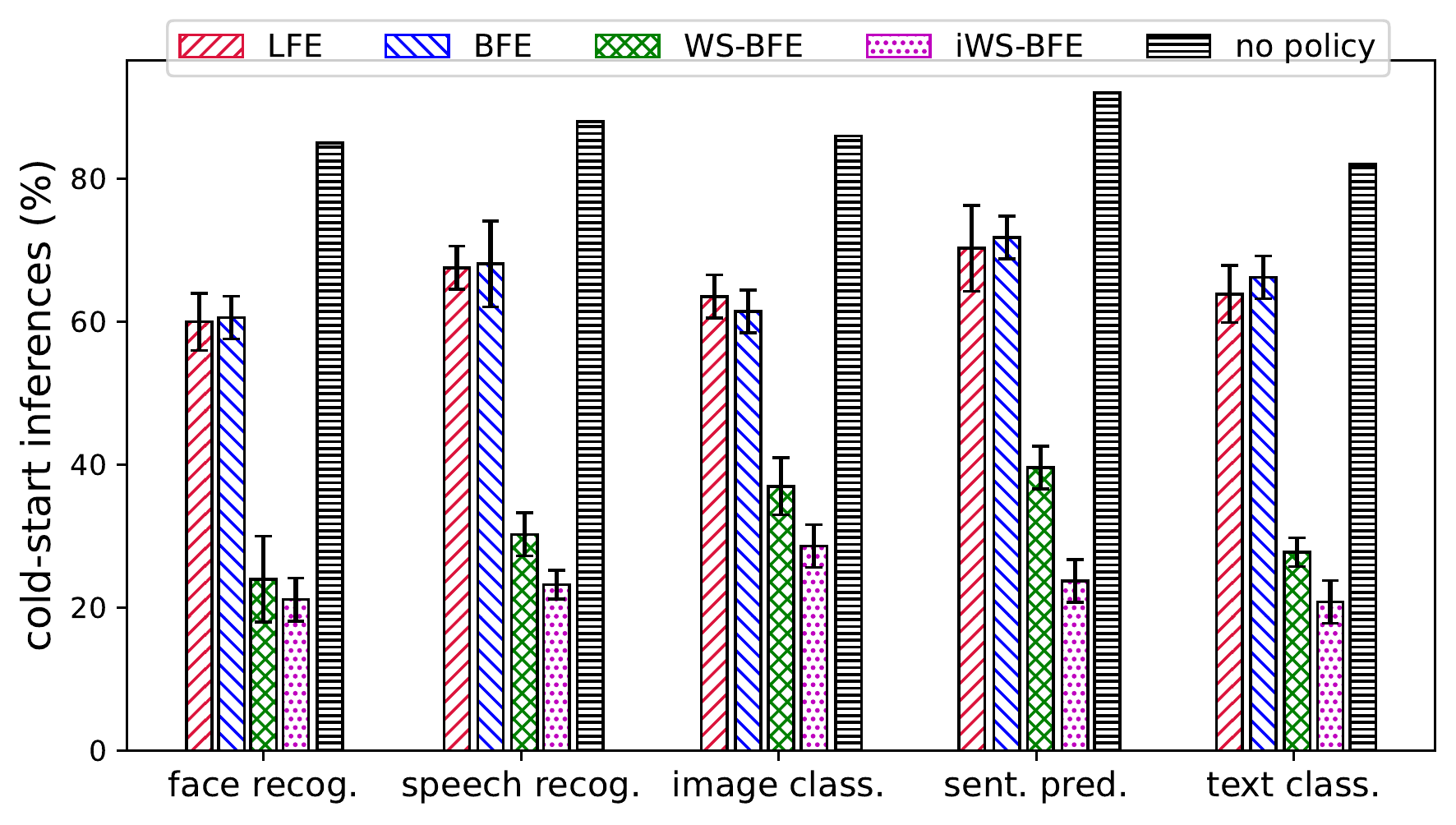}
    \caption{The percentage of cold-start inferences using different NN model eviction policies versus no policy.}
\label{fig:appcold}
%\end{wrapfigure}
\end{figure}

\begin{figure}
\vspace{-3 pt}
    \centering
    \includegraphics[width=.44\textwidth]{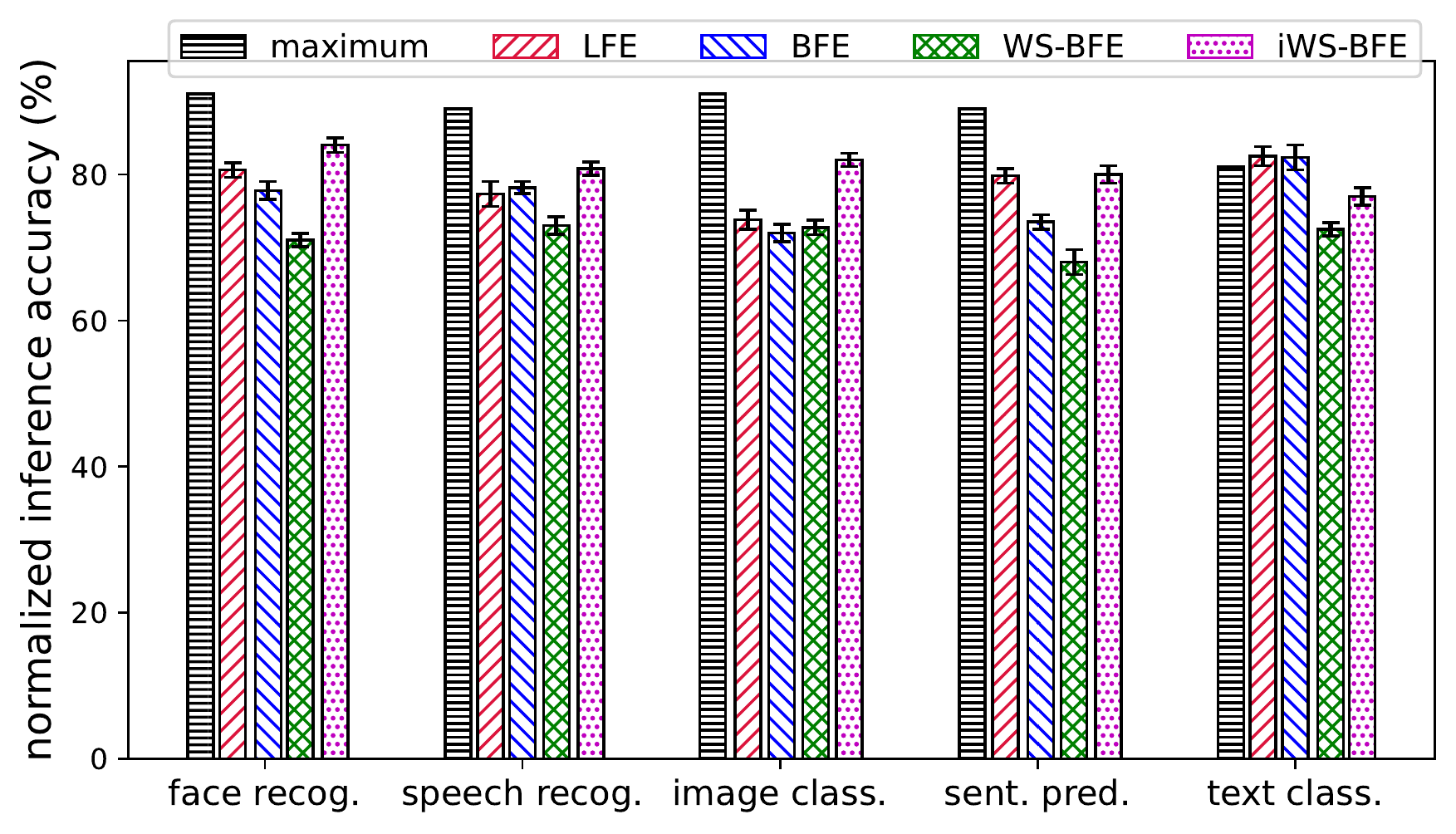}
    \caption{The inference accuracy obtained from the different policies. The ``maximum'' is the benchmark, showing the accuracy of the highest-precision model for each application.}
\label{fig:appaccuracy}
%\end{wrapfigure}
\end{figure}

In this experiment, our goal is to examine whether the achievements of \name and its policies, explored in the previous experiments,  is fairly distributed across all applications, or some applications benefit more than the others. To that end, we analyze the distribution of cold-start inference and accuracy across different DL applications. The name and the NN model characteristics of the examined DL applications are listed in Table~\ref{tab:app_model_size}. Figures~\ref{fig:appcold} and~\ref{fig:appaccuracy}, respectively, express the percentage of cold-start inferences and inference accuracy for each application upon using various NN model eviction policies. It is noteworthy that in Figure~\ref{fig:appcold}, ``no policy'' indicates the situation where \name is not in place, and in Figure~\ref{fig:appaccuracy}, ``maximum'' serve as the benchmark, by showing the use of highest-precision NN model for each application. While Figure~\ref{fig:appcold} shows that WS-BFE and iWS-BFE remarkably outperform the other policies across all the applications, Figure~\ref{fig:appaccuracy} illustrates that, particularly for iWS-BFE, the outperformance does not come with the cost of lower inference accuracy for the applications. More importantly, in both figures, we observe that, for each policy, the percentage of cold-start inferences and accuracy do not fluctuate significantly from one application to the other. This shows that policies are not biased to any particular DL application. Specifically, the rate of cold-start inferences and the accuracy are fairly distributed across different applications.

% Note that, LFE and BFE perform warm-start of the complex models and apparently, applications provide higher accuracy upon adopting any of them. However, warm-starting of large models creates more cold start and even fails to perform execution requests. Due to execution failure in some cases, the mean accuracy of all executions of a particular application is impacted. On the contrary, WS-BFE and iWS-BFE often keep smaller models in warm-start and hence, the mean accuracy is also affected. Although non-replacing policies (LFE, BFE) and warm-start aware (WS-BFE, iWS-BFE) have different trade-offs, their performance is not so distinguishable in this experiment.

%\begin{wrapfigure}{r}{0.75\textwidth}

%% file: Sources/sec-conclusion.tex
\section{Conclusion and Future Work}
\label{sec:conclsn}
Continuous execution of latency-sensitive DL applications is a pressing need of the memory-limited edge systems. The research aims to stimulate the degree of multi-tenancy of such applications without compromising their latency and accuracy objectives.
We developed a framework, called \name, to facilitate multi-tenancy of DL applications via enabling swapping only their NN models. The framework was also equipped with model management policies, particularly iWS-BFE, to choose suitable models for eviction and loading to edge memory, such that the percentage of warm-start inferences is maximized without any major loss in the inference accuracy of the applications. Evaluation results indicate that \name can improve the degree of multi-tenancy by $2\times$, and iWS-BFE can increase warm-start inferences by 60\%. They also show how different policies are robust against uncertainty in the inference request predictions. Last but not the least, the experiments show that the policies are not biased to a certain application in their decisions. There are several avenues to improve \name. One interesting avenue is to partition the models across edge-cloud continuum to reduce their memory footprint on the edge, hence, further improving the degree of multi-tenancy. Another avenue is to use reinforcement learning to enhance the performance of the policies.
%Another avenue is to incorporate other factors such as the energy consumption to improve the model management policies.    
\section{Acknowledgement}
This research is supported by the National Science Foundation (NSF) under awards\# CNS-2007209 and CNS-2117785.